# LMMSE Processing for Cell-free Massive MIMO with Radio Stripes and MRC Fronthaul

Zhifeng Yuan, Yihua Ma, Guanghui Yu

*Abstract*—Cell-free massive MIMO provides ubiquitous connectivity for multiple users, and implementation using radio stripes is very efficient. Compared with collocated massive MIMO, the major cost includes fronthaul overheads and AP hardware. Maximum ratio combination (MRC) achieves a low fronthaul loading and low-cost AP, but the performance is bad. This letter proposes to implement a quasi-LMMSE (Q-LMMSE) processing using MRC fronthaul design. Q-LMMSE is derived from a standard LMMSE, which gains interference information from MRC signal via singular value decomposition. Simulation results show that the proposed Q-LMMSE increases the spectral efficiency by several times using same MRC fronthaul.

## I. INTRODUCTION

Massive multi-input multi-output (MIMO) gains a high spectral efficiency and is proved to be a successful technology for 5G. In future communications like 6G [1], the key performance indicators are increased by a lot. To gain better performance than existing massive MIMO technologies, novel massive MIMO schemes were proposed. Among them, cell-free massive MIMO [2] is very promising, which also brings a trend of shifting from cell-centric network to user-centric network.

Classical massive MIMO employs a collocated architecture, where a large number of antennas are located at base station (BS). The advantage of this scheme is low deployment cost, while the disadvantage is large path loss for far-end users. As a comparison, cell-free massive MIMO [2] employs a distributed architecture, where massive access points (APs) are deployed at different positions. Each AP have single or several local antennas, as well as a simple processing unit. In the uplink, AP simply processes the received signal and then delivers it to central processing unit (CPU). CPU finishes the final processing and demodulation to obtain data information. Cell-free massive MIMO gains marco-diversity against path loss and provides uniformly good coverage for different users.

The gain of cell-free massive MIMO is not free. The cost includes long fronthaul cables and extra AP hardware. Both of them require to be well considered to reduce the cost of practical implementation. One typical fronthaul scheme is that all APs connect to CPU via their separate front-hauls. Maximum ratio combination (MRC) was advocated for its simplicity [2] in early times. User-centric (UC) strategy [3] was applied with MRC to reduce fronthaul loading. With UC method, only a fraction of APs are required for each user, and the performance with MRC is almost not affected. In [4], different levels of cell-free implementations are compared, and a centralized scheme can easily employ linear minimum-mean-square-error (LMMSE) processing to gain high performance. Furthermore, low-resolution analog-to-digital converters (ADCs) was proposed to reduce the cost of cell-free massive MIMO [5]. However, the cell-free scheme using separate fronthauls still suffers from high cost due to lots of long cables.

Another cell-free fronthaul scheme is to use radio stripes [6]. In this scheme, multiple APs share one fronthaul cable for synchronization, data transmission and power supply. By this means, the fronthaul cost is reduced by a lot. It is designed for dense scenarios, e.g., stadiums, stations and malls. A sequential processing [7] named normalized LMMSE (N-LMMSE) was proposed for the cell-free scheme using radio stripes. It makes neighboring APs cooperate to suppress the inter-user interference (IUI), which gains a better performance than MRC. However, AP requires to wait for the results from the previous-stage AP, which brings a large processing latency especially when the number of APs is large. As a comparison, if simple MRC is used, each AP can finish major part of computations in parallel, and the AP processing latency will be much lower. This letter aims at achieving both near-optimal performance and simplest fronthaul/APs for cell-free massive MIMO.

This letter proposes a quasi-LMMSE processing (Q-LMMSE) using MRC fronthaul design. The complexity and latency advantage of MRC fronthaul are still kept, while the performance is improved by a lot. Singular value decomposition (SVD) is used in CPU to gain IUI information from the correlation matrix of MRC signal. LMMSE processing can be approximated using MRC signal and SVD result. The simulation results show that Q-LMMSE gains a much better performance than that of MRC. The major contributions of this letter include: (1) this letter first uses the simple MRC fronthaul to gain a performance much better than MRC, which keeps the complexity and latency advantage; (2) this letter first proposes an SVD-based method to recover IUI information from the correlation matrix of MRC signal; and (3) the proposed Q-LMMSE is very friendly to user-centric and scalable system, which performs even better than centralized LMMSE when same UC is used.

The rest of this letter is organized as follows. In Section II, system model and existing works are introduced. In Section III,

This study is supported by National Key R&D Program of China under Grants 2019YFB1803400.

Y. Ma, Z. Yuan, and G. Yu are with the Department of Wireless Algorithm, ZTE Corporation, Shenzhen, 518057, China; they are also with State Key Laboratory of Mobile Network and Mobile Multimedia Technology, Shenzhen, 518057, China (e-mail: yihua.ma@zte.com.cn; yuan.zhifeng@zte.com.cn; yu.guanghui@zte.com.cn).

Q-LMMSE is proposed and discussed. In Section IV, the simulation results are shown to verify the performance and complexity. Section V briefly concludes this letter. In this letter, $(\cdot)^*$, $(\cdot)^T$, and $(\cdot)^H$ denote conjugate, transpose, and Hermitian transpose, respectively. $\mathbf{I}_N$ is the identity matrix of size $N \times N$, $\mathbb{E}\{\cdot\}$ denote the expectation function, $\|\cdot\|_2$ denotes $l2$ norm and $\oplus$ is the direct sum operation of matrices.

## II. SYSTEM MODEL AND EXISTING WORKS

### A. System Model

A dense scenario is considered as shown in Fig. 1(a). Cell-free massive MIMO with radio stripes is deployed. There are $L$ APs and $K$ active UEs. Each AP has $N$ antennas, so the total antenna number of all APs is $M = NL$. The channel vector of user $k$ is $\mathbf{h}_k = [\mathbf{h}_{k,1}^T, \mathbf{h}_{k,2}^T, ..., \mathbf{h}_{k,L}^T]^T \in \mathbb{C}^M$, where $\mathbf{h}_{k,l} \in \mathbb{C}^N$ is the channel vector between user $k$ and AP $l$. A Rayleigh block fading channel of $\mathbf{h}_{k,l} \sim \mathcal{CN}(\mathbf{0}, \mathbf{R}_{k,l})$ is assumed, where $\mathbf{R}_{k,l} \in \mathbb{C}^{N \times N}$ is the spatial covariance matrix. The large-scale coefficient $\beta_{k,l} = \mathrm{tr}(\mathbf{R}_{k,l})/N$. When $N$ antennas at each AP are uncorrelated, $\mathbf{R}_{k,l} = \beta_{k,l}\mathbf{I}_N$; when $N$ antennas at each AP are correlated, $\mathbf{R}_{k,l}$ is obtained using correlated scattering models [8].

The channel is assumed to be flat over a coherence block of $\tau_c$ channel uses. Among $\tau_c$ channel uses, $\tau_p$ of them are for pilot, while $(\tau_c - \tau_p)$ are for payload data. When $K > \tau_p$, pilot reuse is used. Obviously, pilot reuse leads to pilot contamination. It requires extra pilot allocation to ensure performance, which is not included in this letter and can be found in [9]. An orthogonal pilot set of $\{\boldsymbol{\phi}_1, \boldsymbol{\phi}_2, ..., \boldsymbol{\phi}_{\tau_p}\}$ is used with $\boldsymbol{\phi}_k \in \mathbb{C}^{\tau_p}$ and $\boldsymbol{\phi}_k \in \mathbb{C}^{\tau_p}$. User $k$ is assumed to use pilot indexed by $t_k$, and each use has a set $\mathcal{S}_k = \{i: t_i = t_k\}$. The received signal of pilot $\mathbf{Z}_l \in \mathbb{C}^{N \times \tau_p}$ at AP $l$ is

$$\mathbf{Z}_l = \sum_{i=1}^{K} \sqrt{p_i} \mathbf{h}_{i,l} \boldsymbol{\phi}_{t_i}^T + \mathbf{N}_l \quad (1)$$

where $p_i \geq 0$ is the transmit power, and $\mathbf{N}_l \in \mathbb{C}^{N \times \tau_p}$ is the complex Gaussian noise with independent entries satisfying $\mathcal{CN}(\mathbf{0}, \sigma^2)$. $\sigma^2$ is the average noise power.

Similarly, the received signal of data symbols $\mathbf{Y}_l \in \mathbb{C}^{N \times (\tau_c - \tau_p)}$ at AP $l$ is

$$\mathbf{Y}_l = \sum_{i=1}^{K} \sqrt{p_i} \mathbf{h}_{i,l} \mathbf{s}_i + \mathbf{n}_l \quad (2)$$

where $\mathbf{s}_i \in \mathbb{C}^{1 \times (\tau_c - \tau_p)}$ is the modulated symbols from user $i$ with $\mathbb{E}\{|s_i|^2\} = 1$, and $\mathbf{n}_l \in \mathbb{C}^{N \times (\tau_c - \tau_p)}$ is the complex Gaussian noise with independent entries following $\mathcal{CN}(\mathbf{0}, \sigma^2)$.

### B. Existing Works

The realistic channel estimation (RCE) of [8] $\hat{\mathbf{h}}_{k,l} \in \mathbb{C}^N$ between user $k$ and AP $l$ through LMMSE is

$$\hat{\mathbf{h}}_{k,l} = \sqrt{p_k} \mathbf{R}_{k,l} \left( \sum_{i \in \mathcal{S}_k} \tau_p p_i \mathbf{R}_{i,l} + \sigma^2 \mathbf{I}_N \right)^{-1} \mathbf{Z}_l \boldsymbol{\phi}_{t_k}^* \quad (3)$$

With the channel estimation of each AP, the combination vector $\mathbf{v}_k$ for user $k$ can be obtained to combine the all received signals $\mathbf{y} = [\mathbf{y}_1^T, \mathbf{y}_2^T, ..., \mathbf{y}_L^T]^T \in \mathbb{C}^M$, and the recovered data of user $k$ is $\mathbf{v}_k^H \mathbf{y}$.

Assume that $\hat{\mathbf{h}}_k = [\hat{\mathbf{h}}_{k,1}^T, \hat{\mathbf{h}}_{k,2}^T, ..., \hat{\mathbf{h}}_{k,L}^T]^T \in \mathbb{C}^M$. The combination vector of user $k$ using MRC [2] is $\mathbf{v}_k^{\mathrm{MR}} = \hat{\mathbf{h}}_k$, while that using optimal LMMSE processing [4] is

$$\mathbf{v}_k^{\mathrm{LMMSE}} = \left( \sum_{i \neq k} p_i \left( \hat{\mathbf{h}}_i \hat{\mathbf{h}}_i^H + \mathbf{R}_i - \boldsymbol{\Gamma}_i \right) + \sigma^2 \mathbf{I}_M \right)^{-1} \hat{\mathbf{h}}_k \quad (4)$$

where $(\mathbf{R}_i - \boldsymbol{\Gamma}_i)$ is the channel correlation estimation error matrix. $\mathbf{R}_i \in \mathbb{C}^{M \times M}$ and $\boldsymbol{\Gamma}_i \in \mathbb{C}^{M \times M}$ are

$$\mathbf{R}_i = \mathbf{R}_{i,1} \oplus \mathbf{R}_{i,2} \oplus \cdots \oplus \mathbf{R}_{i,L} \quad (5)$$

$$\boldsymbol{\Gamma}_i = \tau_p p_i \mathbf{R}_i \left( \tau_p p_i \mathbf{R}_i + \sigma^2 \mathbf{I}_M \right)^{-1} \mathbf{R}_i \quad (6)$$

The fronthaul loading of MRC is $2K(\tau_c - \tau_p)$ real numbers, while that of centralized LMMSE processing is $2M(\tau_c - \tau_p) + 2MK$. As $K < M$ in a typical cell-free system, e.g. $M = 96$, $K = 10$ in [7], MRC is able to reduce the fronthaul by a lot. However, MRC performs much worse than LMMSE.

To gain both good performance and low fronthaul loading, N-LMMSE was proposed [7]. The main idea is to compute LMMSE combination of $N$ streams for $K$ users in AP 1, and normalize the noise power of combined streams to $\sigma^2$. Apart from the data estimations of $K$ users, the estimated effective channels and effective channel errors variances are required to transmit to the next stage. For $l > 1$, AP $l$ does an LMMSE to $(N+1)$ received steams including $N$ from its received antennas and one from AP $l$-1. The operations of N-LMMSE are relatively complex and not shown here. Details can be found in [7]. N-LMMSE has both performance and fronthaul loading between MRC and LMMSE.

## III. EFFICIENT PARALLEL SCHEMES

### A. Basic Idea

Although N-LMMSE can improve performance with a relatively low fronthaul loading, it is not efficient in terms of the processing latency. As shown in Fig. 1(b), the computations in a single AP can be divided into two kinds: parallel and serial. The parallel computations can be done simultaneously, while serial computations have to wait for results from the previous AP. As N-LMMSE requires sequential processing, it leads to a large processing latency from AP 1 to CPU. As a comparison, MRC signal $\mathbf{S}^{\mathrm{MRC}} = \mathbf{H}^H \mathbf{Y}$ can be calculated in parallel as

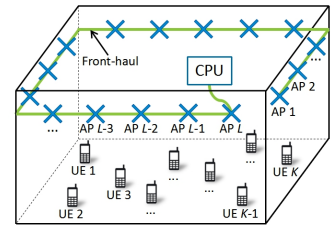

(a) Radio stripe implementation in a large room

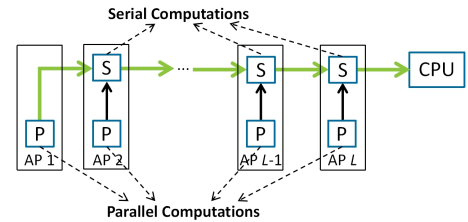

(b) The proposed parallel scheme

Fig. 1 Cell-free massive MIMO using radio stripe implementations. UE represents user equipment.

$$\mathbf{s}_k^{\text{MRC}} = \sum_{l=1}^{L} \mathbf{h}_{k,l}^H \mathbf{Y}_l \tag{7}$$

The multiplications can be done in parallel, and only simple additions are done serially. Also, MRC is very simple, which is friendly to low-cost AP. Therefore, this letter is going to directly reuse the MRC front-haul design to gain a much better performance.

The proposed processing method is derived from LMMSE, and that is why it is named as Q-LMMSE. The standard LMMSE is

$$\mathbf{S}^{\text{LMMSE}} = \mathbf{H}^H \left( \mathbf{H}\mathbf{H}^H + \sigma^2 \mathbf{I}_M \right)^{-1} \mathbf{Y} \tag{8}$$

To make it more related to the MRC signal, an equivalent form [8] is employed as

$$\mathbf{S}^{\text{LMMSE}} = \left( \mathbf{H}^H\mathbf{H} + \sigma^2 \mathbf{I}_K \right)^{-1} \mathbf{H}^H \mathbf{Y} = \left( \mathbf{H}^H\mathbf{H} + \sigma^2 \mathbf{I}_K \right)^{-1} \mathbf{S}^{\text{MRC}} \tag{9}$$

In (9), $\sigma^2$ and $\mathbf{S}^{\text{MRC}}$ can be obtained in MRC fronthaul, while $\mathbf{H}^H\mathbf{H}$ is still needed. As no channel estimation is delivered to CPU in MRC fronthaul design, $\mathbf{H}^H\mathbf{H}$ has to be obtained or estimated from merely MRC data signal.

As $\mathbf{S}^{\text{MRC}} = \mathbf{H}^H \mathbf{Y} = \mathbf{H}^H \mathbf{H}\mathbf{S} + \mathbf{H}^H \mathbf{N}$, its correlation matrix is

$$\mathbf{C}^{\text{MRC}} = \mathbf{S}^{\text{MRC}} \left( \mathbf{S}^{\text{MRC}} \right)^H = L \left( \mathbf{H}^H\mathbf{H}\mathbf{S} + \mathbf{H}^H\mathbf{N} \right)\left( \mathbf{H}^H\mathbf{H}\mathbf{S} + \mathbf{H}^H\mathbf{N} \right)^H \tag{10}$$

Assume that $L_D = \tau_c - \tau_p$. The data signals from different users and noise signal are uncorrelated. When $L_D$ is large enough, $\mathbf{S}\mathbf{S}^H \to L_D \mathbf{I}_L$, $\mathbf{N}\mathbf{N}^H \to M\mathbf{I}_M$ and $\mathbf{S}\mathbf{N}^H \to \mathbf{0}$. Then,

$$\lim_{L_D \to +\infty} \left( \mathbf{C}^{\text{MRC}} / L_D \right) = \left( \mathbf{H}^H\mathbf{H}\mathbf{H}^H\mathbf{H} + \sigma^2 \mathbf{H}^H\mathbf{H} \right) \tag{11}$$

Assume that $\mathbf{A} = \mathbf{H}^H\mathbf{H}\mathbf{H}^H\mathbf{H} + \sigma^2 \mathbf{H}^H\mathbf{H}$. The target matrix $\mathbf{H}^H\mathbf{H}$ can be solved from $\mathbf{A}$ via an SVD-based method described in the next sub-section. For a smaller $L_D$, the approximation of equation (11) is less accurate, which affects the final performance. Different $L_D$ will be used in simulations to verify the robustness of the proposed method.

*B. SVD-based Solution*

As $\mathbf{H}^H\mathbf{H}$ and $\left( \mathbf{H}^H\mathbf{H}\mathbf{H}^H\mathbf{H} + \sigma^2 \mathbf{H}^H\mathbf{H} \right)$ are both Hermitian matrices, they can be unitarily diagonalized. The SVD of $\mathbf{H}^H\mathbf{H}$ is

$$\mathbf{H}^H\mathbf{H} = \mathbf{V}\mathbf{\Lambda}\mathbf{V}^H \tag{12}$$

where $\mathbf{V} \in \mathbb{C}^{K \times K}$ is unitary, and $\mathbf{\Lambda} = \lambda_1 \mathbf{I}_{n_1} \oplus \cdots \oplus \lambda_d \mathbf{I}_{n_d}$ with $\lambda_1 > \lambda_2 > \ldots > \lambda_d \geq 0$ are distinct eigenvalues with respective multiplicities, $n_1, n_2, \ldots, n_d$. In this letter, the eigenvalues in SVD results are assumed to be sorted in order.

Using equation (12), $\mathbf{A}$ becomes

$$\mathbf{A} = \mathbf{V}(\mathbf{\Lambda}\mathbf{\Lambda} + \sigma^2 \mathbf{\Lambda})\mathbf{V}^H = \mathbf{V}\mathbf{\Omega}\mathbf{V}^H \tag{13}$$

where $\mathbf{\Omega} = \left( \lambda_1^2 + \sigma^2 \lambda_1 \right) \mathbf{I}_{n_1} \oplus \cdots \oplus \left( \lambda_d^2 + \sigma^2 \lambda_d \right) \mathbf{I}_{n_d}$ is diagonal. Assume that $\omega_l = \lambda_l^2 + \sigma^2 \lambda_l$ with $l = 1, \ldots, d$, which is also sorted in order like $\lambda_l$. As $\mathbf{V}$ is unitary and $\mathbf{\Omega}$ is diagonal, (13) is an SVD result of $\mathbf{A}$. The eigenvalues of one matrix are unique, while the eigenvectors can be different. Therefore, if SVD is used to $\mathbf{A}$, equation (13) cannot be directly obtained, and the result is assumed to be

$$\mathbf{A} = \mathbf{U}\mathbf{\Omega}\mathbf{U}^H \tag{14}$$

Although (13) and (14) look almost the same, they have different meanings. The main difference is that $\mathbf{V}$ comes from the SVD of $\mathbf{H}^H\mathbf{H}$, while $\mathbf{U}$ comes from the SVD of $\mathbf{A}$. They are different because the eigenvectors of the same matrix is not unique. The target is to express $\mathbf{H}^H\mathbf{H}$ using merely $\mathbf{U}$ and $\mathbf{\Omega}$.

The equation $\omega_l = \lambda_l^2 + \sigma^2 \lambda_l$ is quadratic, and there can be two solutions of $\lambda_l$. The negative solution is eliminated as $\lambda_l > 0$, and the solution of $\lambda_l$ is unique as

$$\lambda_l = \frac{\sqrt{\sigma^4 + 4\omega_l} - \sigma^2}{2} \tag{15}$$

which means $\mathbf{\Lambda}$ in $\mathbf{H}^H\mathbf{H} = \mathbf{V}\mathbf{\Lambda}\mathbf{V}^H$ has been obtained, while $\mathbf{V}$ still requires to be handled. Using (13) and (14),

$$\mathbf{V}\mathbf{\Omega}\mathbf{V}^H = \mathbf{U}\mathbf{\Omega}\mathbf{U}^H \tag{16}$$

Left multiply by $\mathbf{V}^H$ and right multiply by $\mathbf{U}$, (16) becomes

$$\mathbf{\Omega}\mathbf{W} = \mathbf{W}\mathbf{\Omega} \tag{17}$$

where $\mathbf{W} = \mathbf{V}^H\mathbf{U} \in \mathbb{C}^{K \times K}$ is also unitary. According to [10], $\mathbf{W}$ can be expressed as

$$\mathbf{W} = \widetilde{\mathbf{W}}_{n_1} \oplus \cdots \oplus \widetilde{\mathbf{W}}_{n_d} \tag{18}$$

where $\widetilde{\mathbf{W}}_{n_i} \in \mathbb{C}^{n_i \times n_i}$ is unitary, $1 \leq i \leq d$.

From $\mathbf{W} = \mathbf{V}^H\mathbf{U}$, $\mathbf{V} = \mathbf{U}\mathbf{W}^H$. $\mathbf{H}^H\mathbf{H}$ can be written as

$$\mathbf{H}^H\mathbf{H} = \mathbf{V}\mathbf{\Lambda}\mathbf{V}^H = \mathbf{U}\mathbf{W}^H\mathbf{\Lambda}\mathbf{W}\mathbf{U}^H \tag{19}$$

Using (19) and $\mathbf{\Lambda} = \lambda_1 \mathbf{I}_{n_1} \oplus \cdots \oplus \lambda_d \mathbf{I}_{n_d}$,

$$\begin{aligned}
&\mathbf{W}^H \mathbf{\Lambda} \mathbf{W} \\
&= \left( \widetilde{\mathbf{W}}_{n_1} \oplus \cdots \oplus \widetilde{\mathbf{W}}_{n_d} \right)^H \left( \lambda_1 \mathbf{I}_{n_1} \oplus \cdots \oplus \lambda_d \mathbf{I}_{n_d} \right)\left( \widetilde{\mathbf{W}}_{n_1} \oplus \cdots \oplus \widetilde{\mathbf{W}}_{n_d} \right) \\
&= \left( \lambda_1 \widetilde{\mathbf{W}}_{n_1}^H \oplus \cdots \oplus \lambda_d \widetilde{\mathbf{W}}_{n_d}^H \right)\left( \widetilde{\mathbf{W}}_{n_1} \oplus \cdots \oplus \widetilde{\mathbf{W}}_{n_d} \right) \\
&= \left( \lambda_1 \mathbf{I}_{n_1} \oplus \cdots \oplus \lambda_d \mathbf{I}_{n_d} \right) = \mathbf{\Lambda}
\end{aligned} \tag{20}$$

Combing (9), (15), (19) and (20), we can finally have

$$\mathbf{S}^{\text{Q-LMMSE}} = \mathbf{U}^H \left( \mu_1 \mathbf{I}_{n_1} \oplus \cdots \oplus \mu_d \mathbf{I}_{n_d} \right)\mathbf{U}\mathbf{S}^{\text{MRC}} \tag{21}$$

where $\mu_l = \dfrac{2}{\sqrt{\sigma^4 + 4\omega_l} + \sigma^2}$.

Now, Q-LMMSE is obtained using the SVD result of $\mathbf{A}$, which is approximated by MRC fronthaul signal using (11). Another non-ideal factor in practice is channel estimation errors. Ideal channel estimation (ICE) is assumed in the above derivations. However, we only have $\hat{\mathbf{S}}^{\text{MRC}} = \hat{\mathbf{H}}^H \mathbf{Y}$ in practical MRC fronthaul. $\hat{\mathbf{H}}$ is the RCE of $\mathbf{H}$. Even for $L_D \to +\infty$, only $\hat{\mathbf{H}}^H\mathbf{H}\mathbf{H}^H\hat{\mathbf{H}} + \sigma^2 \hat{\mathbf{H}}^H\hat{\mathbf{H}}$ can be obtained from RCE MRC signal in practice. Therefore, even for an enough large $L_D$, Q-LMMSE differs from both ICE LMMSE and RCE LMMSE. It requires a very complex analysis to find out the exact impacts of both $L_D$ and channel estimation errors on the proposed Q-LMMSE. In this letter, we verify these factors via Monte Carlo simulations, which shows that Q-LMMSE is quite robust in various situations and can perform even better than centralized LMMSE in some cases.

IV. SIMULATION RESULTS

*A. Simulation Settings*

RCE is used in all simulations. The simulation scenario is the same as Fig. 1(a). The radio stripe is placed around the top edges of surrounding walls. The length of radio stripe is 800m, and the room size is 200m × 200m × 5m. $L = 24$, $N = 4$, $M = 96$, $\tau_p = K = 24$, and $\tau_c = 720$ or $240$. APs are uniformly distributed at the radio stripe, while users are randomly distributed on the floor.

The correlated channel is generated by Laplace correlated scattering model [8] with an angle spread of 15°. The large-scale fading coefficient (dB) $\beta_{k,l} = -30.5 - 36.7\log_{10}(d_{k,l})$, where $d_{k,l}$ is the distance (m) between user $k$ and AP $l$. The transmit power of each user is 50 mW, and the noise power $\sigma^2 = -92$ dBm. Both correlated and uncorrelated antennas are considered at APs to verify the performance of proposed schemes. When UC is used, each user is detected in only part of APs having the strongest channels for him. Here, the percentage of active APs in UC for each user is set to 1/4.

### B. Performance Comparisons

Fig. 2 shows the cumulative distribution of single user spectral efficiency (SE) for different cell-free schemes. SE of user $k$ is calculated via

$$SE_k = \frac{\tau_c - \tau_p}{\tau_c} E\{\log_2(1 + SINR_k)\} \quad (22)$$

where $SINR_k$ is the receiving SINR of user $k$. A relatively long coherent time $\tau_c = 720$ is used here as a long coherent time is good for the approximation of (11). In the correlated channels scenario, the average SE of Q-LMMSE is 347.1% and 51.3% higher than MRC and N-LMMSE, respectively. SE of Q-LMMSE is close to that of centralized LMMSE. When UC is used, the performances of these schemes apart from MRC are reduced by a lot. One interesting thing is that when UC is used, Q-LMMSE becomes the best method, which is even better than centralized LMMSE. In the scenario of uncorrelated channels, SE performances are improved, compared with the former scenario. The average SE of Q-LMMSE is more than 3 times as large as that of MRC. The performance gap between N-LMMSE and Q-LMMSE decreases, and Q-LMMSE still has 19.7% average SE gain. When UC is used, Q-LMMSE performs best again, and the average SE of Q-LMMSE is 15.4% higher than centralized LMMSE.

In Fig.3, a relatively short coherent time of $\tau_c = 240$ is considered to show that Q-LMMSE also works for general situations. In this scenario, the performance gain of Q-LMMSE becomes smaller. When UC is not used, Q-LMMSE gains a 234.2% average SE advantage over MRC. The average SE of Q-LMMSE is higher than N-LMMSE by 17.5%. When UC is used, Q-LMMSE still achieves the highest SE, which is a 9.9% average SE gain compared with centralized LMMSE. It is also important to notice that if 90%-likely SE is considered, the performance gain of Q-LMMSE becomes even larger as the slope of Q-LMMSE is obvious steeper than others in both Fig. 2 and Fig. 3. 90%-likely SE represents a SE low bound that 90% users can gain.

A heuristic explanation of why Q-LMMSE+UC outperforms centralized LMMSE+UC is shown as follows. In Q-LMMSE, the estimation of $\mathbf{H}^H\mathbf{H}$ comes from $\hat{\mathbf{H}}^H\mathbf{H}\mathbf{H}^H\hat{\mathbf{H}} + \sigma^2\hat{\mathbf{H}}^H\hat{\mathbf{H}}$, which contains the information associated with both ICE and RCE of $\mathbf{H}$. As a comparison, $\hat{\mathbf{H}}^H\hat{\mathbf{H}}$, which contains only the RCE of $\mathbf{H}$, is directly used in centralized RCE LMMSE. This might make Q-LMMSE more stable to channel estimation errors. Also, Q-LMMSE is affected by $L_D$. To visualize the impacts of channel estimation errors and data symbol length, average SE comparison of different ΔSNR and $L_D$ is shown in Fig. 4. ΔSNR defines as a SNR relative to the simulation settings defined in Section 4-A. The change of SNR results from many factors, including transmit power, channel condition and quantization accuracy. When SNR decreases, the gap between the proposed Q-LMMSE and centralized LMMSE becomes smaller. Also, when $L_D$ becomes larger, the performance of Q-LMMSE is relatively better. For the ideal case of $L_D \to +\infty$, Q-LMMSE even performs better than centralized LMMSE at various SNRs. These results suggest that Q-LMMSE is more preferred for the scenarios with relatively large channel estimation errors and long coherent time.

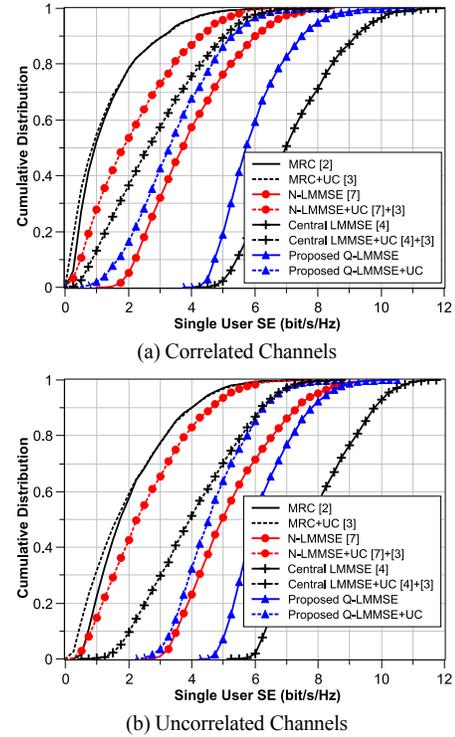

(a) Correlated Channels

(b) Uncorrelated Channels

Fig. 2 Single user SE cumulative distribution comparison of different cell-free schemes for $\tau_c = 720$.

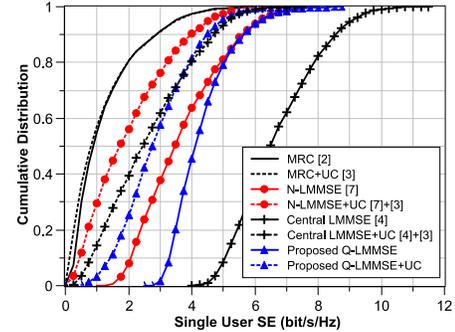

Fig. 3 Single user SE cumulative distribution comparison of different cell-free schemes for correlated channel and $\tau_c = 240$.

Through these performance simulations, two common observations are obtained: (1) when UC is not used, Q-LMMSE gains a better performance than both MRC and N-LMMSE; and (2) when UC is used, Q-LMMSE performs even better than centralized LMMSE. These results show that the proposed Q-LMMSE not only gains a good performance, but also fits for user-centric and scalable cell-free system.

### C. Complexity Comparison

N-LMMSE needs a fronthaul loading of $2K(\tau_c - \tau_p) + 3K^2$ real numbers which has already been lower than that of centralized LMMSE, $2M(\tau_c - \tau_p)$. The proposed Q-LMMSE gain a even lower one, which is only $2K(\tau_c - \tau_p)$. For $\tau_c = 240$, the fronthaul loading of Q-LMMSE is lower than that of N-LMMSE by 1/7. Apart from fronthaul loading, AP complexity and processing latency are also very important in practice. In terms of these two indices, the advantages of Q-LMMSE over N-LMMSE are much greater.

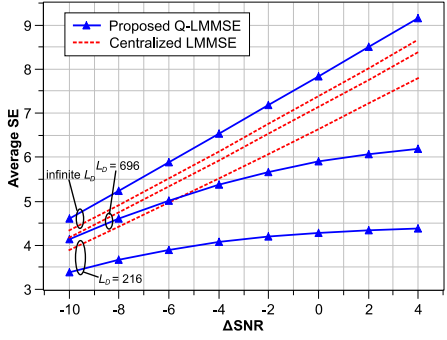

Fig. 4 Average SE comparison of different cell-free schemes for correlated channel and $K = 24$. $\Delta$SNR is a relative value, e.g., when $\Delta$SNR = 0, the simulation settings are the same as those in Fig. 2 and Fig. 3.

The complexity is represented by the number of complex number multiplications. The complexity comparison is shown in Fig. 5(a) and Fig. 5(b). When UC is not used, MRC and Q-LMMSE gain the lowest computational complexity for single AP, which is around half of that of N-LMMSE. The reason is that N-LMMSE require each AP to compute LMMSE of $N+1$ data streams and side information for all users. When UC is applied, all schemes reduce single AP computational complexity by about 75%. With UC, single AP complexity of Q-LMMSE is still 36.8% lower than that of N-LMMSE.

The total time complexity is also important for latency using same AP hardware, or the AP hardware cost for same latency requirement. The total time complexity of all APs is calculated by

$$C_T = L \cdot C_{Serial} + C_{Parallel} \quad (23)$$

where $C_{Serial}$ and $C_{Parallel}$ denote the complexity of calculations in each AP that require to be done serially and can be done in parallel, respectively. As mentioned before, MRC and Q-LMMSE finish major computations in parallel. Only additions are done serially. Therefore, their total time complexity is the same as the maximum complexity of single AP. As a comparison, N-LMMSE requires serial computations, which leads to more than 42 times more total time complexity than the proposed Q-LMMSE. The reason is that only the channel estimation operation can be done in parallel, while other computations require a serial processing. When UC strategy is applied, the parallel complexities are not reduced to exactly 1/4. The reason is that each AP are allocated with different numbers of users to process, and latency is decided by the one which has the most active users. This ratio is obtained from simulations. It shows that UC still reduces the total time complexities of different schemes by around a half. In the UC scenario, the total time complexity of the proposed Q-LMMSE is only around 1/40 as the state-of-art solution, N-LMMSE.

## V. CONCLUSIONS

Cell-free massive MIMO using radio stripes is a very promising B5G/6G technology for dense environment. Key factors of cost and complexity are crucial for the practical implementations. This letter proposes Q-LMMSE to gain both good performance and low-cost cell-free design. Q-LMMSE implements an advanced processing using only simple MRC fronthaul. MRC fronthaul advantages of low fronthaul loading, simple AP hardware and fast parallel computations are kept. The performance is also ensured via approximating LMMSE using only MRC signal. The simulation results also verify the superiority of the proposed Q-LMMSE in terms of both performance and complexity.

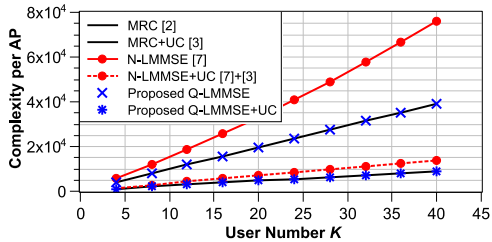
(a) Computational complexity per AP

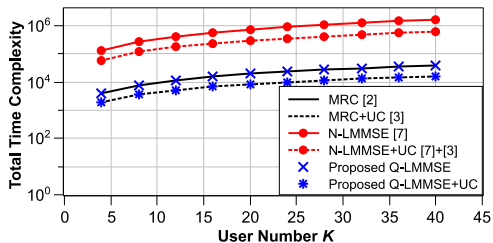
(b) Total time complexity of all AP

Fig. 5 The cost and complexity comparison of different schemes. The common parameters are listed in Section II-A, and $\tau_c = 240$.